\documentclass[aps,showpacs,twocolumn]{revtex4}
\usepackage{amssymb}

\usepackage{graphicx}

\begin{document}

\title{Critical entanglement spectrum of one-dimensional symmetry protected
topological phases}
\author{Wen-Jia Rao$^{1}$, Xin Wan$^{2}$, and Guang-Ming Zhang$^{1,3}$}
\affiliation{$^{1}$State Key Laboratory of Low-Dimensional Quantum Physics and Department
of Physics, Tsinghua University, Beijing 100084, China.\\
$^{2}$Zhejiang Institute of Modern Physics, Zhejiang University, Hangzhou
310027, China\\
$^{3}$Collaborative Innovation Center of Quantum Matter, Beijing, China}
\date{\today }

\begin{abstract}
Under an appropriate symmetric extensive bipartition in a one-dimensional
symmetry protected topological (SPT) phase, a bulk critical entanglement
spectrum can be obtained, resembling the excitation spectrum of the critical
point separating the SPT phase from the trivial (vacuum) state. Such a
critical point is beyond the standard Landau-Ginzburg-Wilson paradigm for
symmetry breaking phase transitions. For the $S=1$ SPT (Haldane) phase with
the Affleck-Kennedy-Lieb-Tasaki exact wave function, the resulting critical
entanglement spectrum has a residual entropy per lattice site $s_{r}=0.67602$%
, showing a delocalized version of the edge excitations in the SPT phase.
From the wave function corresponding to the lowest entanglement energy
level, the central charge of the critical point can be extracted $c\approx
1.01\pm 0.01$. The critical theory can be identified as the same effective
field theory as the spin-1/2 antiferromagnetic Heisenberg chain or the
spin-1/2 Haldane-Shastry model with inverse square long-range interaction.
\end{abstract}

\maketitle

\textit{Introduction}.- Topological properties of low-dimensional quantum
many-body systems have been attracting considerable interest in quantum
information sciences, condensed matter physics and quantum field theories.
It is understood that important information of a topological phase is
encoded in the von Neumann entanglement entropy of its ground state\cite%
{Kitaev-Preskill,Levin-Wen,Zhang-Vishwanath}. In a seminal paper, Li and
Haldane\cite{Li-Haldane} introduced the entanglement spectrum (ES) from the
eigenvalues of the reduced density matrix upon tracing out a subsystem, and
the low-lying part bears a remarkable similarity to the physical edge
spectrum of the topological state\cite%
{cirac-2011,Bernevig-2011,Qi-2012,Pollmann-Oshikawa-2010}. Recently, Hsieh
and Fu\cite{Hsieh-Fu} have suggested that a symmetric extensive bipartition
of the ground state wave function for a topological phase leads to a bulk
critical ES that resembles the excitation spectrum of the critical point
separating the topological phase from the trivial gapped phase. They used an
example of noninteracting fermion Chern insulator to illustrate the emergent
critical ES to describe the phase transition in the integer quantum Hall
effect.

In this Letter, we will examine this proposal by studying a family of
\textit{interacting} topological phases, i.e., symmetry protected
topological (SPT) phases\cite{Chen-Gu-Wen-2011,Schuch,chen-gu-liu-wen}. The
SPT phases possess bulk energy gaps and do not break any symmetry of the
system, but have robust gapless edge excitations. These states can not be
continuously connected to a trivial (vacuum) state without either breaking
the protecting symmetry or closing the energy gap. If the protecting
symmetry is preserved, both SPT phase and the trivial state have the same
symmetry, but differ in the way the protecting symmetry is represented by
their boundary excitations. So there exists a topological phase transition
between the SPT phase and the trivial phase, and the corresponding critical
theory does not belong to the conventional Landau-Ginzburg-Wilson paradigm
for symmetry breaking phase transitions. Such a critical point is dubbed 
"deconfined quantum critical point". Thus, a crucial question is how to
extract the critical properties from the ground state wave function of the
SPT phase.

The simplest example of the SPT phases is the Haldane gapped phase of the
antiferromagnetic spin-1 chain\cite{Haldane-1983}, which is protected by any
one of the following discrete symmetries: time reversal symmetry, link
inversion symmetry, or the $D_{2}\backsimeq Z_{2}\times Z_{2}$ symmetry
comprising $\pi $ rotations about two orthogonal axes\cite%
{Gu-Wen-2009,Pollmann-Oshikawa-2010}. According to the classification
theory, there exists only one nontrivial SPT phase. It is well-known that a
fixed point wave function of the Haldane phase is given by the
Affleck-Kennedy-Lieb-Tasaki (AKLT) valence bond solid state\cite{AKLT}. In
this valence-bond solid picture, the important feature of the Haldane gapped
state is the presence of fractionalized spin-1/2 edge excitations - the
''confined spinons''.

For our purpose to generate a bulk critical spectrum that describes the
continuous phase transition from the nontrivial Haldane phase to a trivial
(vacuum) phase, the deconfined spinons should be emerged as coherent elementary
excitations of the effective critical theory\cite{Babujian}. Hence, we
introduce a symmetric extensive\textit{\ }bipartition and require that the
size of sublattice unit cell is larger than the correlation length of the
AKLT state $\xi \approx 0.91024$. So the simplest choice is to divide the
system into two sublattices both of which have two sites per unit cell. Then
we expect that the critical theory can be characterized by a delocalized
version of the edge spinons of the Haldane phase. This result is supported
by our numerical finding. For a periodic AKLT chain, a critical and
extensive ES with a residual entropy per site is obtained $s_{r}=0.67602$,
compared to $\ln 2$ for the AKLT gapped state under usual left-right
bipartition.

From the wave function corresponding to the lowest entanglement energy
level, we can further calculate the nested entanglement entropy\cite%
{Lou-Katsura,Tanaka}, from which the central charge of the critical point
can be precisely extracted $c\approx 1.01\pm 0.01$. The corresponding
critical theory can be identified as the same effective field theory as the
spin-1/2 antiferromagnetic Heisenberg chain or the Haldane-Shastry spin
model with inverse square long-range interaction\cite{Haldane-Shastry},
namely the (1+1) dimensional SU(2) level-1 Wess-Zumino-Witten field theory.

\textit{Model and method.}- The spin-1 AKLT parent Hamiltonian on a periodic
chain is defined by\cite{AKLT}
\begin{equation}
H=J\sum_{i}\left[ \mathbf{s}_{i}\mathbf{s}_{i+1}+\frac{1}{3}\left( \mathbf{s}%
_{i}\mathbf{s}_{i+1}\right) ^{2}\right] .
\end{equation}%
The corresponding exact ground state wave function for $J>0$ is expressed as%
\begin{equation}
{\small \left| \Psi \right\rangle =}\sum_{\{s_{i}\}}\text{Tr}\left( \mathbf{A%
}^{\left[ s_{1}\right] }\mathbf{A}^{\left[ s_{2}\right] }...\mathbf{A}^{%
\left[ s_{L}\right] }\right) {\small \left| \text{s}_{1},s_{2},...s_{L}%
\right\rangle ,}
\end{equation}%
where $s_{i}=-1,0,+1$ are the local physical spin states and the local
matrices are given by%
\begin{eqnarray}
{\small A^{[-1]}} &{\small =}&{\small \left(
\begin{array}{cc}
0 & 0 \\
-\sqrt{\frac{2}{3}} & 0%
\end{array}%
\right) ,A^{[0]}=\left(
\begin{array}{cc}
-\frac{1}{\sqrt{3}} & 0 \\
0 & \frac{1}{\sqrt{3}}%
\end{array}%
\right) ,}  \nonumber \\
{\small A^{[+1]}} &{\small =}&{\small \left(
\begin{array}{cc}
0 & \sqrt{\frac{2}{3}} \\
0 & 0%
\end{array}%
\right) .}
\end{eqnarray}%
In the thermodynamic limit, the spin-spin correlation function decays
exponentially with a correlation length $\xi =1/\ln 3\simeq 0.91024$, and
any two spins on the even number of the lattice sites are
antiferromagnetically correlated. When we make a cut in real space, the
periodic spin chain transforms into an open chain with fractionalized
spin-1/2 edge spins. For a sufficient long length, these two
edge spins are almost free, leading to four degenerate ground states.
Actually each edge spin contributes two fold degeneracies. Thus, the
degenerate ground state has a residual entropy per edge given by $s_{r}=\ln
2 $.

For a general many-body wave function, the quantum entanglement properties
between the subsystem A and subsystem B are characterized by the reduced
density matrix $\mathbf{\rho }_{A}$, which is formally written as the
''thermal'' density matrix of an entanglement Hamiltonian,%
\begin{equation}
\mathbf{\rho }_{A}=\text{Tr}_{B}|\Psi \rangle \langle \Psi |=e^{-H_{E}},
\end{equation}%
where the wave function $|\Psi \rangle $ is assumed to be normalized. The
full set of eigenvalues of $H_{E}$ denoted by $\{\zeta _{m}\}$ constitutes
the ES of subsystem A. These eigenvalues are directly related to the
coefficients in the Schmidt decomposition of the ground state wave function.
Among all entanglement states in subsystem A, those states with small
eigenvalues have the larger weights in the ground state. The von Neumann
entanglement entropy can be calculated by
\begin{equation}
S=-\text{Tr}\left( \mathbf{\rho }_{A}\ln \mathbf{\rho }_{A}\right)
=\sum_{m}\zeta _{m}e^{-\zeta _{m}}.
\end{equation}

For the AKLT wave function, when the open spin chain is further cut in the
middle of the chain, and two symmetric subsystems A and B represent the left
and right halves of the chain. Then we can easily calculate the ES,
resulting a \textit{single} entanglement energy level $\zeta =\ln 2$ with
two fold degeneracies\cite{rao-zhang-yang}. This implies that there exist
degenerate edge excitations localized around the position where the cut is
made. So such an ES from the left-right bipartition just produces single
physical edge spectrum of the Haldane gapped phase.

In order to reveal the critical properties from the AKLT wave function, we
somehow have to introduce extensive edge modes and make them delocalized
inside the bulk system. So the simplest symmetric extensive bipartition is
proposed and displaying in Fig.1a, where the subsystems A and B are related
by a translation or reflection symmetry. Since the correlation length of the
spin-1 AKLT state is just less than one lattice spacing, the lattice system
is divided into two sublattices A and B, both of which contain two lattice
sites per unit cell.
\begin{figure}[t]
\includegraphics[width=8cm]{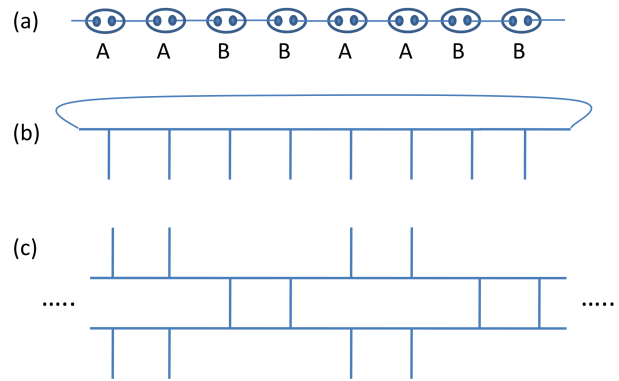}
\caption{(Color online) (a) The symmetric extensive bipartition chosen for
the spin-1 AKLT model Hamiltonian. The small dots are denoted the spin-1/2
degrees of freedom and two spin-1/2 operators form the spin-1 objects. (b)
The matrix product state as the exact ground state wave function. (c) The
reduced density matrix upon tracing out the sublattice B.}
\end{figure}

Consider a finite periodic spin chain with the length $L=4N$. The exact
ground state wave function has the matrix product form and is denoted in
Fig.1b. According to the above symmetric extensive partition, the wave
function basis are regrouped as%
\begin{eqnarray*}
&&{\small \left| \text{s}_{1},s_{2},...s_{L}\right\rangle } \\
&{\small =}&{\small \left| \text{s}_{1}s_{2},s_{5}s_{6},...s_{4N-3}s_{4N-2}%
\right\rangle \left| \text{s}_{3}s_{4},s_{7}s_{8},...s_{4N-1}s_{4N}\right%
\rangle .}
\end{eqnarray*}%
Then we can formally regard $(s_{1}s_{2},s_{5}s_{6},...s_{4N-3}s_{4N-2})$
and $\left( s_{3}s_{4},s_{7}s_{8},...s_{4N-1}s_{4N}\right) $ as two groups
of indices for a large matrix $M=$Tr$\left( \mathbf{A}^{\left[ s_{1}\right] }%
\mathbf{A}^{\left[ s_{2}\right] }...\mathbf{A}^{\left[ s_{4N}\right]
}\right) $, and the wave function is simplified as
\begin{equation}
{\small \left| \Psi \right\rangle }\equiv \sum_{i,j}M_{i,j}|\psi
_{A}^{i}\rangle |\psi _{B}^{j}\rangle ,
\end{equation}%
where $|\psi _{A}^{i}\rangle $ and $|\psi _{B}^{j}\rangle $ represent the
orthonormal basis of the subsystems A and B, respectively. So the ground
state density matrix is expressed as%
\begin{equation}
\mathbf{\rho }=\sum_{i,j}\sum_{m,n}M_{i,j}M_{m,n}^{\ast }|\psi
_{A}^{i}\rangle |\psi _{B}^{j}\rangle \langle \psi _{A}^{m}|\langle \psi
_{B}^{n}|.
\end{equation}%
When all the degrees of freedom of the subsystem B are traced out, the
reduced density matrix is given by%
\begin{equation}
\mathbf{\rho }_{A}=\sum_{i,m}\left( MM^{\dagger }\right) _{i,m}|\psi
_{A}^{i}\rangle \langle \psi _{A}^{m}|,
\end{equation}%
which is represented by Fig.1c. To get the eigenvalues of $\rho _{A}$, it is
more convenient to do singular value decomposition: $M=U\Lambda V^{\dagger }$%
, where $\Lambda $ is a real diagonal matrix. Since $MM^{\dagger }=U\Lambda
V^{\dagger }V\Lambda U^{\dagger }=U\Lambda ^{2}U^{\dagger }$, $\Lambda ^{2}$
gives rise to the eigenvalues of $\rho _{A}$ and hence the entanglement
energy levels $\{\zeta _{m}\}$ are obtained.

\textit{Numerical results.}- The entanglement energy levels $\zeta _{m}$ are
calculated numerically for different length of the chain $%
L=4,8,12,16,20,24,28$. Here we have kept \textit{all} the entanglement
energy states. The corresponding entanglement spectra are displayed in
Fig.2. For each system size, the lowest entanglement energy level $\zeta
_{0} $ is always singlet, corresponding to a nondegenerate ground state,
while the second entanglement level $\zeta _{1}$ is always triplet,
representing the first three-fold excited state. The energy differences
between any two entanglement energy levels decrease as the system size
grows. There is no apparent energy gap in the ES, as indicated in the
largest system size $L=28$. Importantly, we notice that the total number of
entanglement energy levels increases as $2^{L/2}$, suggesting that the
fundamental objects in the reduced subsystem A with a length $L_{r}=L/2$ are
characterized by the spin-1/2 degrees of freedom living on the bonds instead
of the lattice sites.
\begin{figure}[t]
\includegraphics[width=8cm]{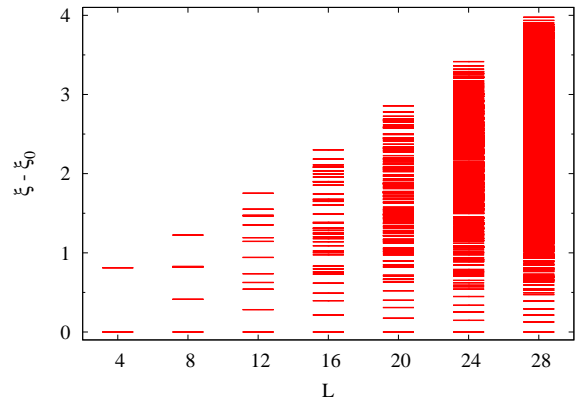}
\caption{(Color online) Entanglement spectrum of the spin-1 AKLT ground
state wave function under a symmetric extended partition for different
length of the chain.}
\label{fig:entanglementSpectrum}
\end{figure}

The lowest entanglement energy level with the largest weight $\xi _{0}$
represents the ground state energy of the entanglement Hamiltonian $H_{E}$.
Although the original AKLT model Hamiltonian only includes the nearest
neighbor interactions, the entanglement Hamiltonian $H_{E}$ generally
involves the long-range interactions. In Fig.3a, the numerical result of $%
\xi _{0}$ is shown as a function of system size $L$. The numerical data can
be fitted by $\xi _{0}=0.27L-0.17$. The constant term can become even
smaller when data of small size systems is neglected. In the thermodynamic
limit, $\xi _{0}$ is expected to depend linearly on the system size,
suggesting that the subsystem A is extensive. Typically, the difference of
the lowest two entanglement energy levels $\xi _{1}-\xi _{0}$ is displayed
as a function of the inverse lattice size in Fig.3b, and it can be fitted by
a power law $L^{-\alpha }$ with $\alpha =0.957\pm 0.003$. Similarly, the
differences of any other two levels vanish in the same limit. Such a
behavior suggests that the resulting ES becomes gapless in the thermodynamic
limit. Thus, the ES from the symmetric extensive bipartition for the AKLT
ground state wave function is extensive and gapless, providing the critical
properties of the topological phase.
\begin{figure}[t]
\includegraphics[width=8cm]{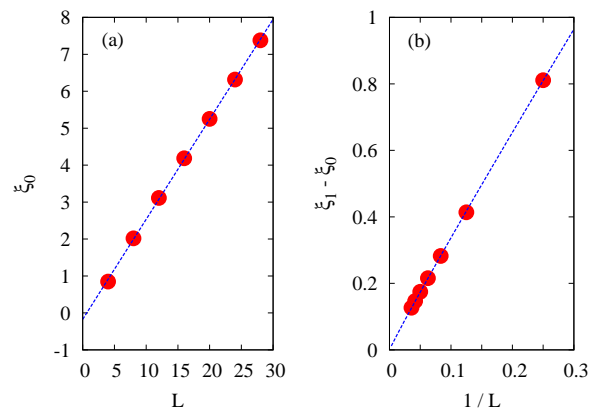}
\caption{(Color online) (a) The lowest entanglement energy level $\protect%
\xi _{0}$ as a function of system size $L$. (b) The difference of the lowest
two entanglement energies $\protect\xi _{1}-\protect\xi _{0}$ as a function
of the inverse system size. }
\label{fig:energyGap}
\end{figure}

To describe this bulk critical ES, we calculate the von Neumann entanglement
entropy per lattice site $s_{r}=S/L_{r}$. The numerical result is shown in
Fig.4. As the system size increases, the entropy quickly saturates at $%
s_{r}=0.676$, within $5\%$ from the free spin-1/2 entropy $\ln 2$. This
seems to suggest that the partitioned subsystem A consists of extensive
delocalized spin-1/2 degrees freedom, namely, the edge spinons of the
Haldane gapped phase is delocalized throughout the bulk system.
\begin{figure}[t]
\includegraphics[width=8cm]{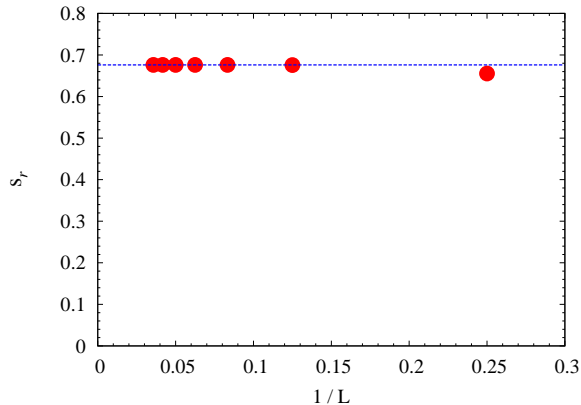}
\caption{(Color online) The von Neumann entanglement entropy per lattice
site. As system size increases, the entropy saturates at $0.676$.}
\label{fig:extendedEntanglementEntropy}
\end{figure}

\textit{Universality class of the critical point.}- In order to determine
the universality class of the critical point from the ES, it is noticed that
the reduced density matrix $\mathbf{\rho }_{A}$ and the entanglement
Hamiltonian $H_{E}$ share the same ''ground state'' wave function:
\begin{equation}
\mathbf{\rho }_{A}|\Psi _{0}\rangle =e^{-\mathbf{\zeta }_{0}}|\Psi
_{0}\rangle ,\text{ }H_{E}|\Psi _{0}\rangle =\zeta _{0}|\Psi _{0}\rangle .
\end{equation}%
Actually, we can further divide the system described by $H_{E}$ with the
length $L_{r}$ into two subsystems, where the length of one subsystem is $l$
and the other is $L_{r}-l$. A so-called nested entanglement matrix can be
introduced as\cite{Lou-Katsura,Tanaka}%
\begin{equation}
\mathbf{\rho }(l)=\text{Tr}_{l+1,l+2,...L_{r}}\left( |\Psi _{0}\rangle
\langle \Psi _{0}|\right) ,
\end{equation}%
where Tr$_{l+1,l+2,...L_{r}}$ means the tracing out over the degrees of
freedom of the subsystem with the length $L_{r}-l$. The nested entanglement
entropy can be defined by
\begin{equation}
s(l,L_{r})=-\text{Tr}_{1,2,...l}\left[ \mathbf{\rho }\left( l\right) \ln
\mathbf{\rho }\left( l\right) \right] ,
\end{equation}%
where the trace is taken over the states in the subsystem with the length $l$%
. At the critical point, according to the conformal field theory\cite%
{Calabrese-Cardy}, the nested entanglement entropy for the systems with
periodic boundary condition is described by%
\begin{equation}
s(l,L_{r})={\frac{c}{3}}\ln \left[ {\frac{L_{r}}{\pi }}\sin \left( \frac{{%
\pi l}}{L_{r}}\right) \right] +s_{0},
\end{equation}%
where $c$ is the central charge of the entanglement Hamiltonian and $s_{0}$
is a non-universal constant. The central charge underpins the conformal
field theory of the critical point.

The numerical result for the nested entanglement entropy $s(l,L_{r})$,
resulting from bipartition of the ground state wave function $|\Psi
_{0}\rangle $ is displayed as a function of $g(l,L_{r})={\frac{L_{r}}{\pi }}%
\sin \left( \frac{{\pi l}}{L_{r}}\right) $ in Fig.5. The data can be well
fitted with the central charge $c=1.01\pm 0.01$. In fact, from the critical
ES, we can further uncover the subset operator content of the critical point
beyond the central charge\cite{Lauchli}. Here we would like to emphasize
that the central charge obtained by the nested entanglement entropy can
determine the critical properties of the entanglement Hamiltonian. The
entanglement Hamiltonian may thus represent the model Hamiltonian of the
critical point of the phase transition between the nontrivial Haldane phase
and the trivial (vacuum) phase. Moreover, we can further find that the
lower-lying entanglement energy part resembles the excitation spectrum of
the low-energy theory for the spin-1/2 antiferromagnetic Heisenberg chain,
more precisely, that for the antiferromagnetic spin-1/2 Haldane-Shastry
model with inverse square long-range interaction\cite{Haldane-Shastry}.
According to the conformal field theory, the effective theory of both models
is characterized by the (1+1) dimensional SU(2) level-1 Wess-Zumino-Witten
conformal field theory\cite{Affleck-Haldane}, where the elementary
excitations are the ''spinon'' particles, satisfying the semion statistics%
\cite{Ludwig}. Actually such a gapless critical theory can also be regarded
as the edge theory of a (2+1) dimensional SPT phase\cite{Liu-Wen}.
\begin{figure}[t]
\includegraphics[width=8cm]{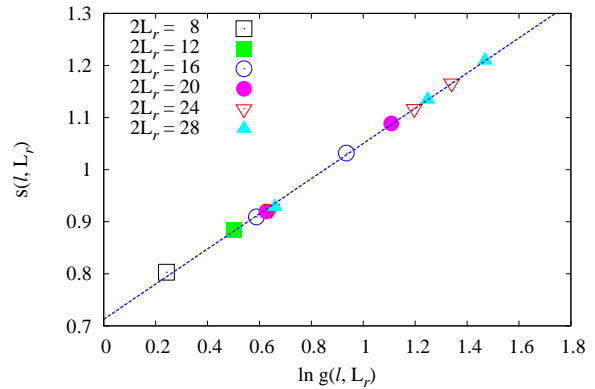}
\caption{(Color online) The nested entanglement entropy $s(l,L_{r})$, as a
function of $g(l,L_{r})$, where $L_{r}$ is the size of the subsystem, while $%
l$ and $L_{r}-l$ are the lengths of the two nested subsystems. }
\label{fig:nestedEntanglementEntropy}
\end{figure}

\textit{Discussion and Conclusion.}- Through the analysis of quantum entanglement
properties with a symmetric extensive bipartition of the AKLT wave function,
the information on the critical point of the topological Haldane phase to
the trivial gapped phase can intriguingly emerge, as long as the size of the
sublattice unit cell is larger than the correlation length of the Haldane
phase. Other types of symmetric extensive bipartitions have also been
considered, leading to the similar results. This further implies that it is
the structure of the edge excitations of the bulk topological phase
determines the minimal size of the sublattice unit cell. Like other
entanglement-based methods, the selection of states according to the
eigenvalues of the reduced density matrix effectively distills the
topological (edge) degree of freedom as the leading contributions, which are
insensitive to the size of the sublattice unit cell. This guarantees the
general applicability of our method to models with larger correlation
length, as long as the edge spectrum of the bulk phase can be described by
a small dimensional space, such as in a SPT phase.

We also found that any \textit{asymmetric} extensive bipartition for the
AKLT wave function always leads to an ES with multiple gaps. So the gapped
excitation spectrum of the Haldane gapped phase can not be obtained through
the asymmetric extensive bipartition. Moreover, we have noticed that the
symmetries of the model Hamiltonian for the SPT phase are relevant in
determining the corresponding effective field theory of the critical point.
Finally an important correspondence between the critical theory and its
gapless boundary theory of the Haldane gapped phase can be established
generally. The boundary theory of the Haldane gapped phase is the critical
theory spatially confined between the topological phase and the trivial
gapped phase\cite{Chen-Wang-Lu-Lee}, while the coherence between extensive
edge excitations (deconfined spinons) emerges as the bulk critical theory
of the topological Haldane phase.

For a general SPT phase in two dimension, when an appropriate symmetric
extensive (checkerboard) bipartition for the ground state wave function is
introduced and satisfy the condition that the characteristic length of the
sublattice unit cell is larger than the correlation length of the
topological SPT phase, a critical ES in the same spatial dimension can be
obtained, and the effective critical field theory of its phase transition to
the trivial gapped phase can be determined. The resulting critical theory is
of topological in nature. Beyond the SPT phases, like Hsieh and Fu\cite%
{Hsieh-Fu} used a single-particle Chern insulator to derive a bulk critical
ES describing the phase transition of the integer quantum Hall affect, a
conceptually similar but richer scenario may also be expected in fractional
quantum Hall systems\cite{Read,Gils}. These related works are under
investigations.

\textit{Acknowledgements.- }GMZ would like to thank D. H. Lee and T. Xiang
for the stimulating discussions and acknowledge the support of NSF-China
through the grant No.20121302227. XW acknowledges the support by the 973
Program under Project No. 2012CB927404 and NSF-China Project No. 11174246.


\begin{thebibliography}{99}
\bibitem{Kitaev-Preskill} A. Kitaev and J. Preskill, Phys. Rev. Lett.
\textbf{96}, 110404 (2006).

\bibitem{Levin-Wen} M. Levin and X. G. Wen, Phys. Rev. Lett. \textbf{96},
110405 (2006).

\bibitem{Zhang-Vishwanath} Y. Zhang, T. Grover, A. Turner, M. Oshikawa, and
A. Vishwanath, Phys. Rev. B \textbf{85}, 235151 (2012).

\bibitem{Li-Haldane} H. Li and F. D. M. Haldane, Phys. Rev. Lett. \textbf{101%
}, 010504 (2008).

\bibitem{cirac-2011} J. I. Cirac, D. Poilblanc, N. Schuch, and F.
Verstraete, Phys. Rev. B \textbf{83}, 245134 (2011).

\bibitem{Bernevig-2011} A. Chandran, M. Hermanns, N. Regnault, and B. A.
Bernevig, Phys. Rev. B \textbf{84}, 205136 (2011).

\bibitem{Qi-2012} X. L. Qi, H. Katsura, and A. W. W. Ludwig, Phys. Rev.
Lett. \textbf{108}, 196402 (2012).

\bibitem{Pollmann-Oshikawa-2010} F. Pollmann, A. M. Turner, E. Berg, and M.
Oshikawa, Phys. Rev. B \textbf{81}, 064439 (2010).

\bibitem{Hsieh-Fu} T. H. Hsieh and L. Fu, arXiv:1305.1949.

\bibitem{Chen-Gu-Wen-2011} X. Chen, Z. C. Gu, and X. G. Wen, Phys. Rev. B
\textbf{83}, 035107 (2011).

\bibitem{Schuch} N. Schuch, D. Perez-Garcia, and I. Cirac, Phys. Rev. B
\textbf{84}, 165139 (2011).

\bibitem{chen-gu-liu-wen} X. Chen, Z. C. Gu, Z. X. Liu, and X. G. Wen, Phys.
Rev. B \textbf{87}, 155114 (2013).

\bibitem{Haldane-1983} F. D. M. Haldane, Phys. Lett. \textbf{93A}, 464
(1983); Phys. Rev. Lett. \textbf{50}, 1153 (1983).

\bibitem{Gu-Wen-2009} Z. C. Gu and X. G. Wen, Phys. Rev. B \textbf{80},
155131 (2009).

\bibitem{AKLT} I. Affleck, T. Kennedy, E. H. Lieb, and H. Tasaki, Phys. Rev.
Lett. \textbf{59}, 799 (1987); Commun. Math. Phys. \textbf{115}, 477 (1988).

\bibitem{Babujian} J. Babujian, Phys. Lett. \textbf{90A}, 479 (1982); Nucl.
Phys. \textbf{B125}, 317 (1983).

\bibitem{Lou-Katsura} J. Lou, S. Tanaka, H. Katsura, N. Kawashima, Phys.
Rev. B \textbf{84}, 245128 (2011).

\bibitem{Tanaka} S. Tanaka, R. Tamura, and H. Katsura, Phys. Rev. A \textbf{%
86}, 032326 (2012).

\bibitem{Haldane-Shastry} F. D. M. Haldane, Phys. Rev. Lett. \textbf{60},
635 (1988); B. S. Shastry, Phys. Rev. Lett. \textbf{60}, 639 (1988).

\bibitem{rao-zhang-yang} W. J. Rao, G. M. Zhang, and K. Yang, Phys. Rev. B
\textbf{89}, 125112 (2014).

\bibitem{Calabrese-Cardy} P. Calabrese and J. Cardy, J. Stat. Mech.: p06002
(2004).

\bibitem{Lauchli} A. M. Lauchli, arXiv:1303.0741.

\bibitem{Affleck-Haldane} I. Affleck and F. D. M. Haldane, Phys. Rev. B
\textbf{36}, 5291 (1987).

\bibitem{Ludwig} P. Schuch and K. Schoutens, Nucl. Phys. B \textbf{482}, 345
(1996).

\bibitem{Liu-Wen} Z. X. Liu and X. G. Wen, Phys. Rev. Lett. \textbf{110},
067205 (2013).

\bibitem{Chen-Wang-Lu-Lee} X. Chen, F. Wang, Y. M. Lu, and D. H. Lee, Nucl.
Phys. B \textbf{873} (FS) 248 (2013).

\bibitem{Read} J. Dubail, N. Read, and E. H. Rezayi, Phys. Rev. B \textbf{86}%
, 245310 (2012).

\bibitem{Gils} C. Gils, E. Ardonne, S. Trebst, A. W. W. Ludwig, M. Troyer,
Z. Wang, Phys. Rev. Lett. \textbf{103}, 070401 (2009).
\end{thebibliography}
\end{document}